# Predicting therapeutic and aggravating drugs for hepatocellular carcinoma based on tissue-specific pathways

Liang Yu[1]*, Fengdan Xu[1], Lin Gao[1]

[1]School of Computer Science and Technology, Xidian University, Xi'an, PR China

Associate Editor: Liang Yu

*Corresponding author

E-mail: lyu@xidian.edu.cn(LY)

## Abstract

**Motivation:** Hepatocellular carcinoma (HCC) is a significant health problem worldwide and annual number of cases are nearly more than 700,000. However, there are few safe and effective therapeutic options for HCC patients. Here, we propose a new approach for predicting therapeutic and aggravating drugs for HCC based on tissue-specific pathways, which considers not only liver tissue and functional information of pathways, but also the changes of single gene in pathways.

**Results:** Firstly, we map genes related to HCC to the liver-specific protein interaction network and get an extended tissue-specific gene set of HCC. Then, based on the extended gene set, 12 enriched KEGG functional pathways are extracted. Using Kolmogorov–Smirnov statistic, we calculate the therapeutic scores of drugs based on the 12 tissue-specific pathways. Finally, after filtering by Comparative Toxicogenomics Database (CTD) benchmark, we get 3 therapeutic drugs and 3 aggravating drugs for HCC. Furthermore, we validate the 6 potentially related drugs of HCC by analyzing their overlaps with drug indications reported in PubMed literatures, and also making CMap profile similarity analysis and KEGG enrichment analysis based on their targets. All analysis results suggest that our approach is effective and accurate for discovering novel therapeutic options for HCC and it can be easily extended to other diseases. More importantly, our method can clearly distinguish therapeutic and aggravating drugs for HCC, which also can be used to indicate unmarked drug-disease associations in CTD as positive or negative.

**Contact:** lyu@xidian.edu.cn(LY)

## 1 Introduction

Hepatocellular carcinoma (HCC), also called malignant hepatoma, is a major health problem being one of the leading causes of death worldwide with annual number of cases exceeding 700,000. It mostly affects patients with liver cirrhosis and currently is their most common reason of death. However, there are few safe and effective therapeutic options for HCC patients. Drug development takes too long and costs too much before taking new drugs to market (Booth and Zemmel, 2004). By rough estimating, it takes about 15 years (DiMasi, 2001) and $800 million to bring a single drug to market (Adams and Brantner, 2006). Drug repositioning has been used as a profitable and successful strategy for decades to save time and money for drug discovery and development. Because the existing drugs are usually vetted in terms of safety, dosage and toxicity, they can be applied in clinical phases more rapidly than newly developed drugs (Ashburn and Thor, 2004). Well-known examples include Minoxidil (originally designed for hypertension; now indicated for hair loss) (Lucky *et al.*, 2004), Sildenafil (originally designed for pulmonary hypertension; now indicated for erectile dysfunction) (Boolell *et al.*, 1996), Amphotericin B (originally designed for serious systemic fungal infections; now indicated for acute promyelocytic leukemia)(Kleinotiene *et al.*, 2013) and Lipitor (originally designed for cardiovascular disease; now indicated for Alzheimer's) (Feldman *et al.*, 2010).

As the development of technology over the past decade, researchers have proposed many computational methods for new drug indications through various strategies. Human genetics studies offer the strongest evidence to connect genes to human diseases. Zhang *et al.* (2016) identified 22 liver cancer-related enhancer SNPs through integrating GWAS data and histone modification ChIP-seq data in 2016. Sanseau *et al.* (2012) found that disease genes extracted from GWAS data were enriched 2.7-fold in drug targets. Besides, they uncovered 123 drug targets which were associated with GWAS traits when not to consider their original indications. These mismatches opened the door for drug repositioning. Bamshad *et al.* (2011) dissected the genetic basis of diseases and traits by exon sequencing which is the targeted sequencing of the subset of the human genome that encodes proteins. Their strategies have focused on identifying individual genes that exhibit differences between normal and disease states. In fact, genes within a cell do not function alone. They interact with each other to carry out specific biological functions. Complex diseases are generally caused by the dysregulation of a set of genes that share common biological function.

Some pathway-based computational approaches were proposed recently to overcome this problem. Ye *et al.* (2012) integrated known drug target information and proposed a disease-oriented strategy for evaluating the relationship between drugs and disease based on their pathway profiles. Pathway profile is a vector with 185 dimensions and each element in the vector is a p-value of the hypergeometric test corresponding to 185 KEGG pathways (Kanehisa *et al.*, 2010). In the next year, Li and Lu (2013) constructed a multi-layer causal network (CauseNet) consisting of chains from drug to target, target to pathway, pathway to downstream gene, and gene to disease. The transition likelihood of each causal chain in the network was measured by a statistical method based on known drug-disease treatment relationships. Zhao *et al.*(2013) extended the concept of cancer signaling bridges (CSB) and developed a computational model to derive specific downstream signaling pathways that revealed previously unknown target-disease connections and new mechanisms for specific cancer subtypes. Pathway analysis can effectively uncover the perturbation of biological systems by diseases. However, a common problem of these methods is that they treat pathways simply as gene sets and ignore the changes of single gene in pathways and tissue-specific information of pathways. Moreover, these methods cannot clearly distinguish therapeutic and aggravating drugs for diseases.

In this study, we propose an approach to identify related drugs for treatments of HCC, which considers the functional information of pathways, the changes of single gene in pathways and the special relationship between HCC and liver tissue. More importantly, our method can clearly distinguish the potential drugs are therapeutic or aggravating for HCC, which can be used to indicate unmarked drug-disease associations in CTD as positive or negative. Fig 1 gives the flowchart of our proposed method. First, we map all the genes related to HCC to the tissue-specific protein interaction network of liver and get an extended tissue-specific gene set of HCC. Then, based on the extended gene set, we extract 12 liver-specific pathways by KEGG pathway enrichment analysis. Using Kolmogorov–Smirnov statistic (Hollander *et al.*, 2013 ), we calculate the therapeutic scores of drugs based on the 12 pathways. Finally, after filtering by Comparative Toxicogenomics Database (CTD) benchmark (Davis *et al.*, 2015), we get top-6 potential drugs according to their scores, in which 3 drugs have the probability to treat HCC and the other three drugs may aggravate HCC. We validate the 6 predicted drugs by analyzing their overlaps with drug indications reported in PubMed literatures. In addition, we also make Connectivity Map (CMap)(Lamb *et al.*, 2006) profiles similarity analysis and KEGG enrichment analysis on their related genes. All these findings suggest that our approach is effective for accurate discovering novel therapeutic options for HCC and easily to be extended to other diseases.

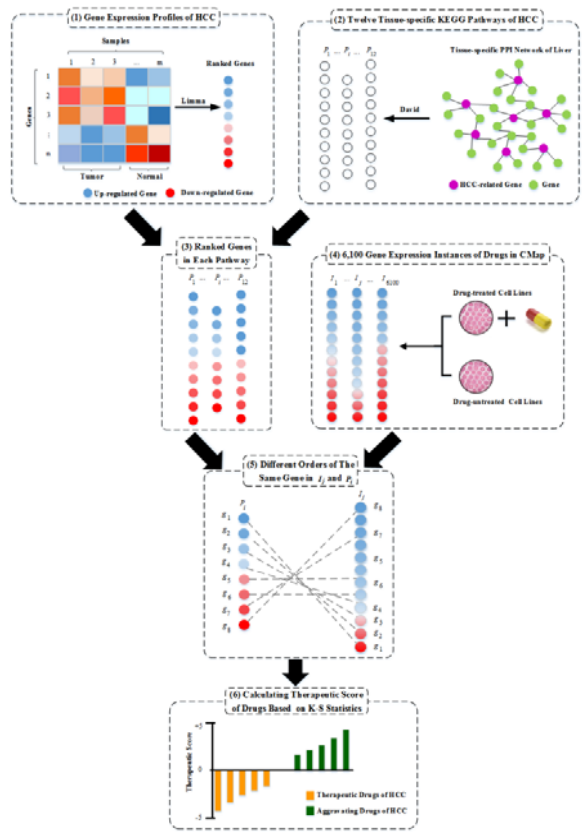

**Fig 1.The framework of our method to predict related drugs for HCC.**
(1) We compare control samples with tumor samples by Limma to generate a list of up-regulated and down-regulated genes for HCC disease. Genes are ranked by their expression values. The red circles represent up-regulated genes, while the blue circles represent down-regulated genes. (2) We integrate disease related genes and liver tissue-specific network to construct a tissue-specific genes signature of HCC. Based on the tissue-specific gene sets, we extract 12 related pathways by KEGG pathway enrichment analysis. The purple circles represent the related genes of HCC, the other green circles represent other genes in the network. (3) Genes in each pathway are ranked based on their expressions obtained from (1). (4) 6100 gene expression instances of drugs from the Connectivity Map. (5) Different orders of the same gene in pathway $P_i(i=1,2,...,12)$ and instance $I_j(j=1,2,...,6100)$. Based on the different orders of genes in pathway $P_i(i=1,2,...,12)$ and instance $I_j(j=1,2,...,6100)$, we calculate the therapeutic score of each drug using Kolmogorov-Smirnov statistic. (6) Drugs are ranked according to their therapeutic scores. The orange bars (therapeutic scores < 0) represent the therapeutic drugs and the green bars (therapeutic scores > 0) represent the aggravating drugs.

## 2 Methods

### 2.1 Extract tissue-specific subnetwork of HCC.

HCC is a cancer that starts in the liver. In order to obtain the tissue-specific subnetwork of HCC, we first get nine related genes of HCC from Online Mendelian Inheritance in Man (OMIM) (Hamosh *et al.*, 2005) , shown in Table 1. OMIM is a comprehensive, authoritative compendium of human genes and genetic disorders to support research and education in human genomics and the practice of clinical genetics. Then, the nine genes related to HCC are extended through obtaining the direct neighbors of nine genes in liver-specific protein-protein interaction (PPI) network got from GIANT(Greene *et al.*, 2015). GIANT is a dynamic, interactive web interface of tissue-specific networks. Finally, we obtain a subnetwork from the liver PPI network, which contains 58 genes and 838 edges with *weight* ≥ 0.1 (S2 Table). This subnetwork has high average clustering coefficient and node degree, and small average shortest path length. The detail information is shown in Table 2. That is to say, the subnetwork has a considerable density and the correlations between 9 HCC genes with other genes are very strong. Hence, we take the 58 genes in the subnetwork as the new gene set related to HCC.

**Table 1.** HCC related genes extracted from OMIM

| Gene Names | Gene Entrez IDs |
| --- | --- |
| IGF2R | 3482 |
| CASP8 | 841 |
| MET | 4233 |
| PDGFRL | 5157 |
| TP53 | 7157 |
| PIK3CA | 5290 |
| LCO | 3935 |
| CTNNB1 | 1499 |
| AXIN1 | 8312 |

**Table 2.** Network topological attributes of the tissue-specific sub-network of HCC

| Attribute names | Values |
| --- | --- |
| Number of nodes | 58 |
| Number of edges | 838 |
| Average node degree | 28.9 |
| Average clustering coefficient | 0.69 |
| Average the shortest path length | 1.5 |
| Maximal diameter | 3 |

### 2.2 Construct HCC tissue-specific pathway signatures.

Because HCC is a complex disease and complex diseases generally involve several gene mutations and pathway dysregulations (Chen *et al.*, 2009; Chen *et al.*, 2010), we further analyze the biological functions of 58 genes got from the tissue-specific PPI network of HCC. By using DAVID tool (Huang *et al.*, 2009), we perform pathway enrichment analysis on these genes. The parameters of DAVID are set as: p-value = 0.001 and count = 5. Finally, we obtain 12 KEGG pathways (Kanehisa *et al.*, 2010) related to the 58 genes (see Table 3). There are totally 28 of 58 (48.3%) tissue-specific genes are enriched in these 12 KEGG pathways. Therefore, we name the 12 KEGG pathways as HCC-specific pathways (shown in Fig 1(2)).

Through analyzing the gene expression data of HCC (see section "Datasets"), we can get a ranked gene list of HCC based on their expression values. The genes in the list are divided into two sets: the up-regulated set and the down-regulated set (shown in Fig 1(1)). We compare control samples with tumor samples by Limma (Smyth, 2004) to generate a list of up-regulated and down-regulated genes for HCC disease. Genes are ranked by $\log FC$, actually $\log FC = \log_2(\text{mean}(\text{tumor})/\text{mean}(\text{control}))$, where $\text{mean}(\text{tumor})$ and $\text{mean}(\text{control})$ represent the average of the expression value of a gene across all tumor samples and all control samples respectively. If the $\log FC$ of a gene is greater than zero, then the gene is up-regulated, otherwise the gene is down-regulated. Then, based on the ranked gene list, we rank the genes in each HCC-specific pathway (shown in Fig 1 (3)). In this way, we obtain 12 ranked HCC-specific gene lists, which are named as HCC tissue-specific pathway signatures (S3 Table3).

**Table 3.** Twelve enriched KEGG pathways with HCC

| Pathways | Number of HCC-specific genes | P-values |
| --- | --- | --- |
| Pathways in cancer | 20 | 3.06E-13 |
| Prostate cancer | 10 | 1.41E-08 |
| Adherens junction | 8 | 1.44E-06 |
| Endometrial cancer | 7 | 2.15E-06 |
| Colorectal cancer | 8 | 2.62E-06 |
| Apoptosis | 8 | 3.32E-06 |
| Melanoma | 7 | 1.36E-05 |
| Wnt signaling pathway | 9 | 1.41E-05 |
| Cell cycle | 7 | 3.28E-04 |
| Notch signaling pathway | 5 | 4.16E-04 |
| Basal cell carcinoma | 5 | 7.61E-04 |
| Melanogenesis | 6 | 8.61E-04 |

### 2.3 Calculate the therapeutic scores of drugs.

We get 6100 gene expression instances covered 1,309 drugs from Cmap database. In other words, a drug may correspond to multiple instances. We rank the genes in each instance by taking the differential expression values between drug-treated and drug-untreated cell lines (shown in Fig 1(4)). Finally, we get 6100 drug-related gene lists. If the gene expression profiles related to a drug positively correlate to that of HCC-specific pathways, the drug may worsen HCC. On the contrary, if their relationship is negative, the drug may treat HCC. The reason is that the positive correlations show that drugs may strengthen the expressions of disease genes and the negative correlations represent that the drugs may inhibit the expressions of disease genes. Therefore, based on 12 HCC-specific pathways and 6100 drug-related gene expression instances, we use a nonparametric, rank-based pattern matching strategy originally introduced by Lamb et al. (2006) to evaluate the relationship between drugs and HCC (shown in Fig 1 (5)).

We take each ranked drug expression profile as reference signature and assess their similarity to each HCC pathway. For a HCC pathway $i$ and a drug expression profile $j$, we compute a connectivity score separately for the set of up- or down-regulated genes in the pathway $i$: $CS_{up}^{i,j}$ or $CS_{down}^{i,j}$.

For the HCC pathway $i$, it needs up- ($G_{up}$) or down-regulated genes ($G_{down}$) as inputs. The computational formulas as follows (Lamb *et al.*, 2006):

$$a = \underset{p=1}{\overset{m}{Max}} \left[ \frac{p}{m} - \frac{V(p)}{n} \right] \quad (1)$$

$$b = \underset{p=1}{\overset{m}{Max}} \left[ \frac{V(p)}{n} - \frac{p-1}{m} \right] \quad (2)$$

$$CS = \begin{cases} a \text{ (if a > b)} \\ -b \text{ (if a < b)} \end{cases} \quad (3)$$

Where $n$ represents the total number of genes in the reference drug expression profile $j$; $m$ represents the number of up-regulated genes $G_{up}$, or the number of down-regulated genes $G_{down}$; $p$ represents the po-sition of the input set ( $p = 1...m$ ); $V(p)$ is the position of the $p$th input gene in the gene list of drug expression profile $j$; $CS$ represents the connectivity score $CS_{up}^{i,j}$ or $CS_{down}^{i,j}$ for the up- or down-regulated gene set.

The therapeutic score ( $TS$ ) between a drug with $k$ instances in CMap and HCC with 12 functional pathways is calculated as follows:

$$TS = \frac{1}{k}\sum_{i=1}^{12}\sum_{j=1}^{k} CS^{i,j} \qquad (4)$$

where the drug has $k$ instances in CMap and HCC has 12 functional pathways; $CS^{i,j}$ represents the connectivity score between HCC pathway $i$ and instance $j$, which is defined as:

$$CS^{i,j} = CS_{up}^{i,j} - CS_{down}^{i,j} \qquad (5)$$

where $CS_{up}^{i,j}$ and $CS_{down}^{i,j}$ represent the connectivity scores for the up-regulated gene set and the down-regulated gene of pathway $i$ respectively, and their definitions are given in formula (3) to (5).

If the up-regulated pathway genes are near the top (up-regulated) of the rank-ordered drug gene lists and the down-regulated pathway genes are near the bottom (down-regulated) of the rank-ordered drug gene lists, we can get high positive therapeutic scores (TS), which indicate the drugs and the HCC pathways have similar expression profiles and the drugs might aggravate HCC. On the other hand, if the up-regulated pathway genes are near the bottom of the rank-ordered drug gene lists and the down-regulated pathway genes are near the top of the rank-ordered drug gene lists, we can get negative therapeutic scores(TS), which imply the given drugs and the HCC pathways have adverse expression profiles and the drugs could be treatment candidates for HCC. Both cases are significant for HCC treatment.

### 2.4 Datasets

HCC gene expression data. We download the gene expression profiles of HCC from the NCBI Gene Expression Omnibus (GEO)(Edgar *et al.*, 2002). The gene expression dataset (GSE45436) contains 134 samples, including 95 tumor samples and 39 control samples. Each probe is mapped to a gene using GPL570 platform obtained from GEO, while the probes are removed if they do not match any gene. If multiple microarray probes are mapped to a same gene, we assign the average of expression values to the gene.

Here, the expression values of all genes are standardized as follows:

$$z_{ij} = \frac{g_{ij} - mean(g_i)}{std(g_i)} \qquad (6)$$

where $g_{ij}$ represents the expression value of gene $i$ in sample $j$ , and $mean(g_i)$ and $std(g_i)$ respectively represent mean and standard deviation of the expression vector for gene $i$ across all samples.

Gene expression data related to drugs. The gene expression data related to drugs is downloaded from the CMap database (Lamb *et al.*, 2006). It contains 6,100 instances, which cover 1,309 drugs. These instances are measured on five types of human cancer cell lines, including the breast cancer epithelial cell lines MCF7 and ssMCF7, the prostate cancer epithelial cell line PC3, the nonepithelial lines HL60 (leukemia) and SKMEL5 (melanoma).

## 3 Results

### 3.1 Choose Potential HCC Drugs through CTD Benchmark

The quick brown fox jumps over the lazy dog. The quick brown fox jumps over the lazy dog. The quick brown fox jumps over the lazy dog. The quick brown fox jumps over the lazy dog. The quick brown fox jumps over the lazy dog. The quick brown fox jumps over the lazy dog. The quick brown fox jumps over the lazy dog. The quick brown fox jumps over the lazy dog. The quick brown fox jumps over the lazy dog. The quick brown fox jumps over the lazy dog. The quick brown fox jumps over the lazy dog. The quick brown fox jumps over the lazy dog. The quick brown fox jumps over the lazy dog.

To find most likely HCC-related drugs, we need evaluate the precision of our method firstly. We take CTD (Davis *et al.*, 2015) as benchmark. CTD provides manually curated information about drug-gene interactions, drug-disease and gene-disease relationships. Curated chemical-disease associations are extracted from the published literature by CTD biocurators and inferred associations are established via CTD curated chemical-gene interactions.

For a drug in CMap, if it cannot find corresponding chemical name in CTD, we will not calculate its therapeutic score (defined in section "Methods"). In this way, we finally get 1168 scored drugs. Because most drug-disease associations in CTD are not marked as positive or negative, we rank the 1168 drugs in descending order by the absolute values of their therapeutic scores (S1 Table). We know the top drugs imply stronger connections with HCC. And then we calculate the precisions of our approach at different top-x drugs, which are shown in Fig 2. The precision is calculated as follows:

$$precision = \frac{P_{CTD}}{P} \qquad (7)$$

where $P$ represents the number of top-x drugs; $P_{CTD}$ represents the number of drugs in the top-x drugs can be found related with HCC in CTD database.

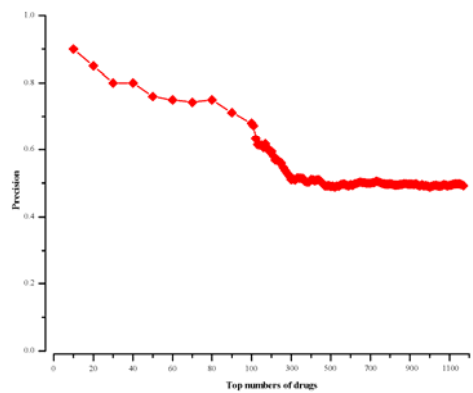

**Fig 2. The precision of our approach at different top-x drugs.**

We find in the top-10 drugs (x=10), there are 9 drugs associated with HCC in CTD. That is to say, the precision is 0.9. For the top-20 drugs (x=20), the precision is 0.85 and there are three potentially HCC-related drugs. When x is 30, its precision is 0.8 and we get six potential drugs with HCC. From the Fig 2, we notice that with the increase of x, the precision declines and the number of potential drugs increases. We hope we can predict relatively more HCC-related drugs with high precision. Therefore, we choose top-30 (x = 30) drugs for further analysis.

## 3.2 Validate potentially HCC-related Drugs through PubMed Literature

In the above section, we choose the top-30 drugs (precision = 0.8) for further analysis. There are 12 therapeutic drugs with negative TS values in the top-30 drugs, shown in Table 4. Nine of them can be found having connections with HCC in CTD. But only one of the 9 drugs is marked as therapeutic drug (Rank=3 and Evidence = "T" in Table 4) for HCC and the other 8 inferred drugs are unmarked in CTD. Here, we can indicate these 8 unmarked drugs are possibly therapeutic drugs for HCC. The rest three drugs (Securinine, Isocarboxazid and Ajmaline) are newly predicted ones by our method, which are marked as bold in Table 4. Based on PubMed, we analyze the three drugs further. PubMed (pubmed.gov) is a free resource developed and maintained by the National Center for Biotechnology Information (NCBI) at the National Library of Medicine (NLM). PubMed comprises more than 26 million references and abstracts on life sciences and biomedical topics.

**Table 4.** Twelve Therapeutic Drugs for HCC in the Top-30 Drugs.

| Rank | Drug Name | Evidence | Ref Count |
|---|---|---|---|
| 1 | Geldanamycin | inferred | 33 |
| 2 | Lomustine | inferred | 1 |
| 3 | **Securinine** | NULL | NULL |
| 4 | Clozapine | inferred | 35 |
| 5 | Suramin | inferred | 14 |
| 6 | Rosiglitazone | inferred | 79 |
| 7 | Furazolidone | inferred | 7 |
| 8 | Hydrocortisone | inferred | 30 |
| 9 | **Isocarboxazid** | NULL | NULL |
| 10 | **Ajmaline** | NULL | NULL |
| 11 | 1-(5-Isoquinolinesulfonyl)-2-Methylpiperazine | inferred | 15 |
| 12 | Troglitazone | T | 86 |

Evidence represents a drug-disease association is curated, inferred or not existed in CTD database. Curated associations include three types: marker/mechanism (Evidence = "M"), therapeutic (Evidence = "T"), marker/mechanism & therapeutic (Evidence = "M&T"). If an association is inferred by CTD, Evidence = "inferred", and if it is not existed in CTD, Evidence = "NULL"; Ref Count represents the number of reference (s) for the curated and inferred associations. If an association is not existed in CTD, Ref Count = "NULL".

Securinine (Rank=3 in Table 4), a quinolizine pseudoalkaloid (not from amino acid) from securinega suffurutiosa or securinini nitras, is one of central nervous stimulants and clinically applied to treat amyotrophic lateral sclerosis (ALS), poliomyelitis and multiple sclerosis (Buravtseva, 1958; Copperman *et al.*, 1973; Copperman *et al.* 1974). It has been found to be active as a γ-amino butyric acid (GABA) receptor antagonist (Beutler *et al.*, 1985). GABA is the chief inhibitory neurotransmitter in the central nervous system and plays a principal role in reducing neuronal excitability throughout the nervous system. Studies show that GABA stimulates HCC cell line HepG2 growth (Li *et al.* 2012). Consequently, it means that securinine is a promising agent with therapeutic potential for HCC through inhibiting GABA receptor. The mechanism is depicted in Fig 3A.

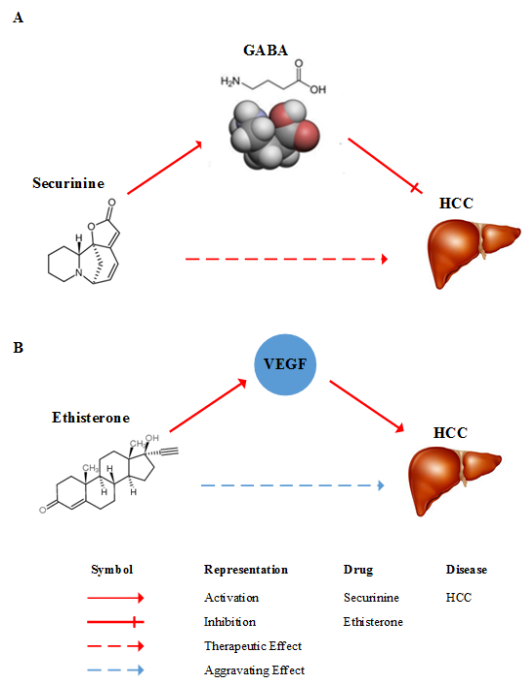

**Fig 3. Description of the mechanism of drugs and HCC.** A. The mechanism of securinine treating HCC. Securinine has been found to be active as a γ-amino butyric acid (GABA) receptor antagonist. GABA stimulates HCC cell line HepG2 growth. Consequently, it means that securinine is a promising agent with therapeutic effect on HCC patients through inhibiting GABA receptor. B. The mechanism of ethisterone aggravating HCC. Ethisterone is a synthesized progestin and it can increase the expression of vascular permeability factor (VEGF). The overexpression of VEGF can contribute to the proliferation of HCC tumor cells. Hence ethisterone gets aggravating effect on HCC patients through increasing the expression level of VEGF.

Isocarboxazid (Rank=9 in Table 4) is a non-selective, irreversible monoamine oxidase inhibitor (MAOI) used as an antidepressant (Fagervall *et al.*, 1986). It has been investigated in the treatment of Parkinson's disease and other dementia-related disorders. MAOIs inhibit the activity of the monoamine oxidase enzyme family and decrease dopamine production (Goldstein *et al.*, 2016). Dopamine is an organic chemical belong to catecholamine and phenethylamine families and plays several important roles in the brain and body. It is synthesized from L-DOPA, which is produced in the brain and kidneys. Studies have revealed that dopamine receptors (DRs) are over expressed in the HCC (Villa *et al.*, 2005; Lu *et al.*, 2015), and the overexpression of DRs improves the chance combining with dopamine. However, isocarboxazid can decrease this chance by reducing dopamine. Therefore, isocarboxazid has the potential role in the treatment of HCC.

Ajmaline (Rank=10 in Table 4) is a sodium ion channel blocker. It has served as a treatment for Wolff-Parkinson-White syndrome (WPW) which is a disorder of the electrical system of the heart referred to arrhythmia with the ventricles contracting prematurely. Sodium channel blockers are drugs which impair conduction of sodium ions (Na+) through sodium channels. The sustained low extracellular sodium ion concentration decrease glutamate uptake (Fujisawa *et al.*, 2015). Glutamine is an essential compound in cellular metabolism and a fuel for cell proliferation, so the decrease of glutamine will inhibit tumor proliferation (Hensley *et al.*, 2013). Hence, ajmaline provides an accessible therapeutic window for HCC treatment.

The other 18 drugs with negative TS values are shown in Table 5. They are possible to aggravate HCC. 15 of them have been found having relationships with HCC in CTD database and we can infer these relationships are possibly negative. The remaining 3 drugs (Pridinol, Piperacillin and Ethisterone) are newly potential drugs for aggravating HCC marked as bold in Table 5. We will investigate the three drugs (Pridinol, Piperacillin and Ethisterone) based on PubMed.

Pridinol (Rank=7 in Table 5) is a muscle relaxant that is able to alleviate pain (Merck, 1989). In PubMed, there are only eight articles related to pridinol. Due to lack of information about pridinol, it is hard for us to find evidences to infer that pridinol may aggravate HCC. We are confident that our prediction for pridinol aggravating HCC will be verified in the future.

Piperacillin (Rank=9 in Table 5) is a well-known β-lactam antibiotic (Holten, 2000). β-lactam antibiotics work by inhibiting the biosynthesis of the peptidoglycan layer of bacterial cell walls. The peptidoglycan layer is the outermost and fundamental component of cell walls, and it is crucial for cell wall structural integrity. Blocking the synthesis of the peptidoglycan layer destroys the bacterial cell wall and induces bacterial cell rupture. Lipopolysaccharides (LPS) are released because of bacterial cell rupture (Parija, 2014). Studies demonstrate that LPS promotes proliferation, adhesion and invasion of hepatoma cells (Liu *et al.*, 2014). Therefore, piperacillin probably aggravates HCC indirectly by distracting the bacterial cell wall.

Ethisterone (Rank=13 in Table 5) is a synthesized progestin. Studies revealed that natural and synthetic progestin increased the expression of vascular permeability factor(VEGF). VEGF is a signal protein involved in vasculogenesis and angiogenesis and its principal function is to produce new blood vessels during embryonic development after blood vessels injury. VEGF also is a potent angiogenesis growth factor. The overexpression of VEGF and its receptors has been associated with poor prognosis in HCC (Dhar *et al.*, 2001; Tseng *et al.*, 2008). Moreover, the overexpression of VEGF can contribute to the proliferation of HCC tumor cells. If HCC patients orally take ethisterone, the level of VEGF will increase. The situation of those patients will be more terrible. Hence, ethisterone gets some evil influence over HCC patients (The mechanism depicts in Fig 3B).

**Table 5.** Eighteen Aggravating Drugs for HCC in the Top-30 Drugs.

| Rank | Drug Name | Evidence | References_count |
|---|---|---|---|
| 1 | Scriptaid | inferred | 7 |
| 2 | Entinostat | inferred | 37 |
| 3 | Trichostatin A | inferred | 89 |
| 4 | Harman | inferred | 1 |
| 5 | Estradiol | inferred | 116 |
| 6 | Riboflavin | inferred | 18 |
| 7 | Pridinol | NULL | NULL |
| 8 | Diethylstilbestrol | inferred | 69 |
| 9 | Piperacillin | NULL | NULL |
| 10 | Nilutamide | inferred | 2 |
| 11 | N-(2-cyclohexyloxy-4-nitrophenyl)-methanesulfonamide | inferred | 32 |
| 12 | Trioxsalen | inferred | 1 |
| 13 | Ethisterone | NULL | NULL |
| 14 | Fulvestrant | inferred | 59 |
| 15 | Acacetin | inferred | 1 |
| 16 | Cetirizine | inferred | 2 |
| 17 | Dinoprost | inferred | 18 |
| 18 | Haloperidol | inferred | 33 |

Evidence represents a drug-disease association is curated, inferred or not existed in CTD database. Curated associations include three types: marker/mechanism (Evidence = "M"), therapeutic (Evidence = "T"), marker/mechanism & therapeutic (Evidence = "M&T"). If an association is inferred by CTD, Evidence = "inferred", and if it is not existed in CTD, Evidence = "NULL"; Ref Count represents the number of reference (s) for the curated and inferred associations. If an association is not existed in CTD, Ref Count = "NULL".

### 3.3 Analyze potentially HCC-related Drugs through CMap database

The CMap database can not only be applied to calculate drug-disease correlations, but also can be used to identify connections between two drugs. In particular, for a same disease, if two drugs have strongly positive relationship, they may have similar effects on this disease. On the contrary, if their relationship is negative, they may have opposite effects. In this section, we further analyze the six predicted drugs (three therapeutic drugs shown in Table 4: Securinine, Isocarboxazid and Ajmaline; three aggravating drugs shown in Table 5: Pridinol, Piperacillin and Ethisterone) based on CMap and estimate their correlations (evaluated by formula (5)) with known HCC drugs marked as "therapeutic" in CTD database. The results are shown in Table 6.

For the three potentially therapeutic drugs (Securinine, Isocarboxazid and Ajmaline) marked as bold in Table 6, we find that securinine yields highly positive connectivity score (calculated by formula (5)) for daunorubicin, troglitazone and paclitaxel. Daunorubicin is a chemotherapy medication used to treat cancer. The liposomal formulation of the anthracycline daunorubicin has low systemic toxicity and is taken up strongly by the liver. The researchers found that daunorubicin was an active agent against HCC (Daniele *et al.*, 1999). Troglitazone is a thiazolidinedione PPARγ agonist that exhibits anti-diabetic, anticancer, anti-fibrotic, and anti-inflammatory activities. Studies (Yu *et al.*, 2006) found a significant decrease in expression of PPARγ mRNA and protein in human liver cancers compared with surrounding nontumorous liver. Troglitazone induces PPARγ expression and inhibits cell growth both in vitro and in vivo (Yu *et al.*, 2006). Paclitaxel is a cancer medication that interferes with the growth and spread of cancer cells in the body (Huang *et al.*, 2015). Studies indicate that paclitaxel is cytotoxic to cultured hepatocellular carcinoma cells (Gagandeep *et al.*, 1999) and is a common chemodrug for therapy of HCC patients (Yan *et al.*, 2013; Chen *et al.*, 2015; Chao *et al.*, 1998).

Isocarboxazid is found having strongly positive relationships with doxorubicin, sirolimus and troglitazone. Doxorubicin has been regarded as one of the effective chemotherapy medication used to treat cancer (Weiss, 1992). One way that doxorubicin works is by blocking an enzyme called topo isomerase 2 that cancer cells need to divide and grow. Sorafenib, a multikinase inhibitor, has been shown to be effective and safe monotherapy in patients with advanced HCC. A recent phase II study revealed the safety and efficacy of the combined use of sorafenib and doxorubicin-TACE in patients from the Asia-Pacific region with intermediate HCC (Chung *et al.*, 2013). The disease control rate and overall response rate were reported to be 91.2% and 52.4%, respectively. Also, doxorubicin is used to improve the survival rate of unresectable HCC patients (Shang *et al.*, 2016). Sirolimus is a macrolide compound and is indicated for the prevention of organ transplant rejection. The data suggested a beneficial effect of sirolimus immunosuppression on the recurrence of hepatocellular carcinoma after liver transplantation (Hanish and Knechtle, 2011; Chinnakotla *et al.*, 2009). Moreover, researchers found sirolimus could inhibit the growth of human hepatoma cells (Schumacher *et al.*, 2002).

Ajmaline gets high positive connectivity scores with nimesulide, dexamethasone, doxorubicin and paclitaxel. Nimesulide is a cyclooxygenase-2 (COX-2) inhibitor and is used to inhibit the proliferation and promote apoptosis of Hep G2 (a human liver cancer cell line) by up-regulating Smad4 (Yang *et al.*, 2012). Dexamethasone, a type of corticosteroid medication, can restore gluconeogenesis in malignant hepatocytes by bypassing the abnormal regulation of 11β-HSD enzymes, which indicates that targeting altered metabolism in liver cancer could prove useful as a therapeutic strategy for HCC (Ma *et al.*, 2013).

For the three potentially aggravating drugs (Pridinol, Piperacillin and Ethisterone) in Table 3, they all have negative relationship with known HCC drugs. Pridinol has high negative connectivity scores with genistein, nimesulide, troglitazone and sirolimus. Piperacillin have clear negative connection scores with dexamethasone, genistein, paclitaxel and sirolimus. Ethisterone have clear negative connection scores with dexamethasone, genistein, sirolimus and troglitazone. Genistein is one of the most abundant isoflavones in soy that is described as an angiogenesis inhibitor and a phytoestrogen (Walter, 1941). Researcher found genistein and its inhibition of multiple signal transduction pathways could control the invasiveness and metastatic potential of HCC (Wang *et al.*, 2014) and the combination of ATO with genistein would present a promising therapeutic approach for the treatment of HCC (Ma *et al.*, 2011). More details about the other known HCC drugs have been introduced in the above sections.

**Table 6.** The relationships of six predicted drugs with known HCC therapeutic drugs in CTD.

| Predicted Drugs | Known HCC Drugs in CTD | Connectivity Scores |
|---|---|---|
| **Securinine** | Daunorubicin | 0.397 |
|  | Troglitazone | 0.504 |
|  | Paclitaxel | 0.321 |
| **Isocarboxazid** | Doxorubicin | 0.31 |
|  | Sirolimus | 0.432 |
|  | Troglitazone | 0.436 |
| **Ajmaline** | Nimesulide | 0.542 |
|  | Dexamethasone | 0.41 |
|  | Doxorubicin | 0.36 |
|  | Paclitaxel | 0.424 |
| **Pridinol** | Genistein | -0.377 |
|  | Nimesulide | -0.269 |
|  | Troglitazone | -0.442 |
|  | Sirolimus | -0.382 |
| **Piperacillin** | Dexamethasone | -0.436 |
|  | Genistein | -0.336 |
|  | Paclitaxel | -0.299 |
|  | Sirolimus | -0.408 |
| **Ethisterone** | Dexamethasone | -0.369 |
|  | Genistein | -0.223 |
|  | Sirolimus | -0.236 |
|  | Troglitazone | -0.267 |

The potentially therapeutic drugs of HCC are marked as bold. The other three drugs are potentially aggravating drugs of HCC.

All in all, these results indicate that our method can find both therapeutic drugs and aggravating drugs for diseases. Moreover, our method can be applied to mark the known drug-disease associations as positive or negative, such as some drug-disease associations in CTD database, to provide a better basis for the treatment of diseases.

### 3.4 Pathway Enrichment Analysis

Furthermore, we try to apply DAVID functional annotation tool (Huang *et al.*, 2009) to identify enriched KEGG pathways of the above six drugs based on their targets. If the targets of a drug have close correlations with HCC-related pathways, it is very likely that the drug is helpful for the treatment of HCC. The pathway details are shown in Table 7. Because securinine, pridinol and ethisterone have no target information at present in DrugBank, they have no related KEGG pathways.

The potentially therapeutic drugs of HCC are marked as bold. The other three drugs are potentially aggravating drugs of HCC. "NULL" represents the drug has no targets in DrugBank at present. Thus, its corresponding KEGG pathway is "NULL" too.

Isocarboxazid has two targets "MAOA and MAOB" and they relate with 13 KEGG pathways. Some of them have been reported having close relationships with liver cirrhosis or HCC. For example, for the alcoholism pathway, the researchers found chronic alcohol consumption had long been associated with progressive liver disease toward the development of hepatic cirrhosis and the subsequent increased risk for developing HCC (McKillop and Schrum, 2009). Metabolic pathways are linked series of chemical reactions occurring within a cell. The results showed that the metabolism in the HCC tumor was modified to promote cellular proliferation or escape from apoptosis (Huang *et al.*, 2013). For the drug metabolism-cytochrome P450 pathway, cytochrome P450 1A1 is a major enzyme in the bioactivation of exogenous procarcinogens of hepatocellular carcinoma (HCC) (Li *et al.*, 2009). For the drug ajmaline, it has one enriched pathway, adrenergic signaling in cardiomyocytes. The adrenergic signaling pathway plays an important role in cancer progression by regulating multiple cellular processes (Cole and Sood, 2012; Fitzgerald, 2012; Fitzgerald, 2009).

**Table 7.** Pathway enrichment analysis result of six selected drugs.

| Drug Name | Drug Targets | KEGG Pathways |
|---|---|---|
| Securinine | NULL | NULL |
| Isocarboxazid | MAOA, MAOB | Phenylalanine metabolism; Histidine metabolism; Tyrosine metabolism; Glycine, serine and threonine metabolism; Tryptophan metabolism; Cocaine addiction; Arginine and proline metabolism; Amphetamine addiction; Drug metabolism - cytochrome P450; Serotonergic synapse; Dopaminergic synapse; Alcoholism; Metabolic pathways |
| Ajmaline | SCN5A | Adrenergic signaling in cardiomyocytes |
| Pridinol | NULL | NULL |
| Piperacillin | PBP3, PENA, PBP2A, PBP1B | Peptidoglycan biosynthesis |
| Ethisterone | NULL | NULL |

There are nine of thirteen pathways in which isocarboxazid drug targets exist and these nine pathways have common genes with the 12 tissue-specific KEGG pathway of HCC (Table 3). The interaction between the nine pathways and the 12 tissue-specific KEGG pathway of HCC is shown in

Fig.4. Tyrosine metabolism has common genes with one pathways which are tissue-specific KEGG pathway of HCC. There are common genes between Cocaine addiction and seven HCC related pathways. For Arginine and proline metabolism, it exits common genes with one HCC related pathway. There are common genes between Amphetamine addiction and seven HCC related pathways. Drug metabolism-cytochrome P450 has common genes with two HCC pathways. Serotonergic synapse has common genes with ten pathways of liver cancer. Dopaminergic synapse has common genes with nine pathways of liver cancer. There are common genes between Alcoholism and five HCC related pathways. For Metabolic pathways, it exits common genes with seven HCC related pathway.

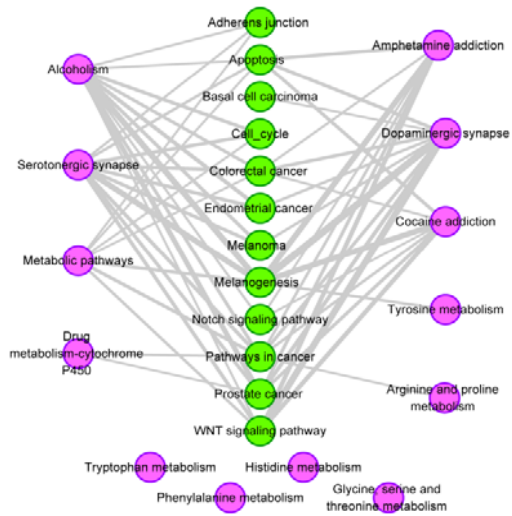

**Figure 4. .Pathway analysis of isocarboxazid drug targets.** The purple circle represents pathways of the mercaptopurine drug targets appeared, and the green circle represents tissue-specific KEGG pathway of HCC. The gray side indicates that there is a common gene between two pathways, and the more number of public genes is, the more rough edge is.

There is only one pathway in which Ajmaline drug targets exist. However, Adrenergic signaling in cardiomyocytes has common genes with ten of 12 tissue-specific KEGG pathway of HCC (Table 3). The interaction between the six pathways and the 12 tissue-specific KEGG pathways of HCC is shown in Fig.5.

These results further show that these potential drugs have close relationships with HCC. With the improvement of data, our algorithm will perform better.

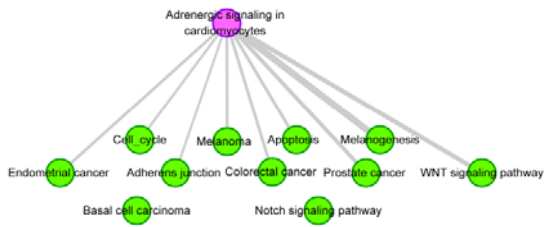

**Figure 5. Pathway analysis of reserpine drug targets.** The purple circle represents pathways of the reserpine drug targets appeared, and the green circle represents tissue-specific KEGG pathway of HCC. The gray side indicates that there is a common gene between two pathways, and the more number of public genes is, the more rough edge is**.**

## 4 Conclusions

Hepatocellular carcinoma (HCC), also called malignant hepatoma, is the most common type of liver cancer. In this study, we propose an approach to identify potential drugs for treatments of HCC, which takes into account the special relationships among HCC, liver tissue and KEGG pathways. First, we map genes related to HCC to the tissue-specific protein interaction network of liver and get an extended gene set of HCC. Then, based on the new tissue-specific gene set of HCC, we extract 12 enriched functional pathways by KEGG pathway enrichment analysis. The 12 KEGG pathways are used to calculate the therapeutic score ($TS$) of each drug using Kolmogorov–Smirnov statistic. High positive therapeutic scores ($TS$) indicate the drugs and the HCC pathways have similar expression profiles and the drugs might aggravate HCC. On the contrary, negative therapeutic scores ($TS$) imply the given drugs and the HCC pathways have adverse expression profiles and the drugs could be treatment candidates for HCC. That is to say, our method can clearly distinguish therapeutic and aggravating drugs for HCC, which make it possible for providing a more accurate reference for the treatment of HCC. Furthermore, it is helpful in marking some known associations as therapeutic or aggravating between drugs and diseases in CTD database. Finally, we choose top-six potentially related drugs of HCC for further analysis, including three therapeutic drugs and three aggravating drugs. They are validated by analyzing their overlaps with drug indications reported in CTD benchmark and PubMed literature. Furthermore, CMap profiles similarity analysis and KEGG enrichment analysis are also made on their related genes. These findings suggest that our approach is more useful and effective in predicting novel therapeutic options for HCC, and it can be easily extended to other diseases. The performance of our proposed method can be further improved by integrating more pathway information and tissue data, which will help predicting novel drugs for diseases in a more accurate way.

## References

. di Villa Bianca, R. D. E., Sorrentino, R., Roviezzo, F., Imbimbo, C., Palmieri, A., De Dominicis, G., ... & Mirone, V. (2005). Peripheral relaxant activity of apomorphine and of a D1 selective receptor agonist on human corpus cavernosum strips. International journal of impotence research, 17(2), 127-133.

Adams, C. P., & Brantner, V. V. (2006). Estimating the cost of new drug development: is it really $802 million?. Health affairs, 25(2), 420-428.

Ashburn, T. T., & Thor, K. B. (2004). Drug repositioning: identifying and developing new uses for existing drugs. Nature reviews Drug discovery, 3(8), 673-683.

Bamshad, M. J., Ng, S. B., Bigham, A. W., Tabor, H. K., Emond, M. J., Nickerson, D. A., & Shendure, J. (2011). Exome sequencing as a tool for Mendelian disease gene discovery. Nature Reviews Genetics, 12(11), 745-755.

Beutler, J. A., Karbon, E. W., Brubaker, A. N., Malik, R., Curtis, D. R., & Enna, S. J. (1985). Securinine alkaloids: a new class of GABA receptor antagonist. Brain research, 330(1), 135-140.

Boolell, M., Gepi‐Attee, S., Gingell, J. C., & Allen, M. J. (1996). Sildenafil, a novel effective oral therapy for male erectile dysfunction. British journal of urology, 78(2), 257-261.

Booth, B., & Zemmel, R. (2004). Prospects for productivity. Nature Reviews Drug Discovery, 3(5), 451-456.

Buravtseva, G. R. (1958). Result of application of securinine in acute poliomyelitis. Farmakologiia i toksikologiia, 21(5), 7-12.

Chao Y; Chan WK; Birkhofer MJ; Hu OY; Wang SS; Huang YS; Liu M; Whang-Peng J; Chi KH; Lui WY; Lee SD. (1998). Phase ii and pharmacokinetic study of paclitaxel therapy for unresectable hepatocellular carcinoma patients. British Journal of Cancer, 78(1), 34-9.

Chen, L., Liu, Y., Wang, W., & Liu, K. (2015). Effect of integrin receptor-targeted liposomal paclitaxel for hepatocellular carcinoma targeting and therapy. Oncology Letters, 10(1), 77.

Chen, L., Wang, R. S., & Zhang, X. S. (2009). Biomolecular networks: methods and applications in systems biology (Vol. 10). John Wiley & Sons.

Chen, L., Wang, R., Li, C., & Aihara, K. (2010). Modeling biomolecular networks in cells: structures and dynamics. Springer Science & Business Media.

Chinnakotla, S., Davis, G., Vasani, S., Kim, P., Tomiyama, K., & Sanchez, E., et al. (2009). Impact of sirolimus on the recurrence of hepatocellular carcinoma after liver transplantation. Liver Transplantation,15(12), 1834–1842.

Chung, Y. H., Han, G., Yoon, J. H., Yang, J., Wang, J., Shao, G. L., ... & Chao, Y. (2013). Interim analysis of START: Study in Asia of the combination of TACE (transcatheter arterial chemoembolization) with sorafenib in patients with hepatocellular carcinoma trial. International journal of cancer, 132(10), 2448-2458.

Cole, S. W., & Sood, A. K. (2012). Molecular pathways: beta-adrenergic signaling in cancer. Clinical cancer research, 18(5), 1201-1206.

Copperman, R., Copperman, G., & Der Marderosian, A. (1973). From Asia securinine--a central nervous stimulant is used in treatment of amytrophic lateral sclerosis. Pennsylvania medicine, 76(1), 36.

Copperman, R., Copperman, G., & Marderosian, A. D. (1974). Letter: securinine. Jama the Journal of the American Medical Association,228(3), 288.

Daniele, B., De Vivo, R., Perrone, F., Lastoria, S., Tambaro, R., Izzo, F., ... & Pignata, S. (1999). Phase I clinical trial of liposomal daunorubicin in hepatocellular carcinoma complicating liver cirrhosis. Anticancer research, 20(2B), 1249-1251.

Davis, A. P., Grondin, C. J., Lennon-Hopkins, K., Saraceni-Richards, C., Sciaky, D., King, B. L., ... & Mattingly, C. J. (2015). The Comparative Toxicogenomics Database's 10th year anniversary: update 2015. Nucleic acids research, 43(D1), D914-D920.

Dhar, D. K., Naora, H., Yamanoi, A., Ono, T., Kohno, H., Otani, H., & Nagasue, N. (2001). Requisite role of VEGF receptors in angiogenesis of hepatocellular carcinoma: a comparison with angiopoietin/Tie pathway. Anticancer research, 22(1A), 379-386.

DiMasi, J. A. (2001). New drug development in the United States from 1963 to 1999. Clinical Pharmacology & Therapeutics, 69(5), 286-296.

Edgar, R., Domrachev, M., & Lash, A. E. (2002). Gene Expression Omnibus: NCBI gene expression and hybridization array data repository. Nucleic acids research, 30(1), 207-210.

Fagervall, I., & Ross, S. B. (1986). Inhibition of mono amine oxidase in monoaminergic neurones in the rat brain by irreversible inhibitors. Biochemical pharmacology, 35(8), 1381-1387.

Feldman, H. H., Doody, R. S., Kivipelto, M., Sparks, D. L., Waters, D. D., Jones, R. W., ... & Breazna, A. (2010). Randomized controlled trial of atorvastatin in mild to moderate Alzheimer disease LEADe. Neurology, 74(12), 956-964.

Fitzgerald, P. J. (2009). Is norepinephrine an etiological factor in some types of cancer?. International journal of cancer, 124(2), 257-263.

Fitzgerald, P. J. (2012). Beta blockers, norepinephrine, and cancer: an epidemiological viewpoint. Clin Epidemiol, 4, 151-156.

Fujisawa, H., Sugimura, Y., Takagi, H., Mizoguchi, H., Takeuchi, H., Izumida, H., ... & Fukumoto, K. (2015). Chronic hyponatremia causes neurologic and psychologic impairments. Journal of the American Society of Nephrology, ASN-2014121196.

Gagandeep, S., Novikoff, P. M., Ott, M., & Gupta, S. (1999). Paclitaxel shows cytotoxic activity in human hepatocellular carcinoma cell lines.Cancer Letters, 136(1), 109.

Goldstein, D. S., Jinsmaa, Y., Sullivan, P., Holmes, C., Kopin, I. J., & Sharabi, Y. (2016). Comparison of monoamine oxidase inhibitors in decreasing production of the autotoxic dopamine metabolite 3, 4-dihydroxyphenylacetaldehyde in PC12 cells. Journal of Pharmacology and Experimental Therapeutics, 356(2), 484-493.

Greene, C. S., Krishnan, A., Wong, A. K., Ricciotti, E., Zelaya, R. A., Himmelstein, D. S., ... & Chasman, D. I. (2015). Understanding multicellular function and disease with human tissue-specific networks. Nature genetics, 47(6), 569-576.

Hamosh, A., Scott, A. F., Amberger, J. S., Bocchini, C. A., & McKusick, V. A. (2005). Online Mendelian Inheritance in Man (OMIM), a knowledgebase of human genes and genetic disorders. Nucleic acids research, 33(suppl 1), D514-D517.

Hanish, S. I., & Knechtle, S. J. (2011). Liver transplantation for the treatment of hepatocellular carcinoma. Oncology, 25(8), 752.

Hensley, C. T., Wasti, A. T., & DeBerardinis, R. J. (2013). Glutamine and cancer: cell biology, physiology, and clinical opportunities. The Journal of clinical investigation, 123(9), 3678-3684.

Hollander, M., Wolfe, D. A., & Chicken, E. (2013). Nonparametric statistical methods. John Wiley & Sons.

Holten, K. B. (2000). Appropriate prescribing of oral beta-lactam antibiotics. American family physician, 62(3).

Huang, D. W., Sherman, B. T., & Lempicki, R. A. (2009). Systematic and integrative analysis of large gene lists using DAVID bioinformatics resources. Nature protocols, 4(1), 44-57.

Huang, Q., Tan, Y., Yin, P., Ye, G., Gao, P., Lu, X., ... & Xu, G. (2013). Metabolic characterization of hepatocellular carcinoma using nontargeted tissue metabolomics. Cancer research, 73(16), 4992-5002.

Huang, X., Qin, J., & Lu, S. (2015). Up-regulation of mir-877 induced by paclitaxel inhibits hepatocellular carcinoma cell proliferation though targeting foxm1. International Journal of Clinical & Experimental Pathology, 8(2), 1515-24.

Kanehisa, M., Goto, S., Furumichi, M., Tanabe, M., & Hirakawa, M. (2010). KEGG for representation and analysis of molecular networks involving diseases and drugs. Nucleic acids research, 38(suppl 1), D355-D360.

Kleinotiene, G., Posiunas, G., Raistenskis, J., Zurauskas, E., Stankeviciene, S., Daugelaviciene, V., & Machaczka, M. (2013). Liposomal amphotericin B and surgery as successful therapy for pulmonary Lichtheimia corymbifera zygomycosis in a pediatric patient with acute promyelocytic leukemia on antifungal prophylaxis with posaconazole. Medical Oncology, 30(1), 433.

Lamb, J., Crawford, E. D., Peck, D., Modell, J. W., Blat, I. C., Wrobel, M. J., ... & Reich, M. (2006). The Connectivity Map: using gene-expression signatures to connect small molecules, genes, and disease. science, 313(5795), 1929-1935.

Li, J., & Lu, Z. (2013). Pathway-based drug repositioning using causal inference. BMC bioinformatics, 14(16), S3.

Li, R., Shugart, Y. Y., Zhou, W., An, Y., Yang, Y., Zhou, Y., ... & Jin, L. (2009). Common genetic variations of the cytochrome P450 1A1 gene and risk of hepatocellular carcinoma in a Chinese population. European Journal of Cancer, 45(7), 1239-1247.

Li, Y. H., Liu, Y., Li, Y. D., Liu, Y. H., Li, F., Ju, Q., ... & Li, G. C. (2012). GABA stimulates human hepatocellular carcinoma growth through overexpressed GABAA receptor theta subunit. World J Gastroenterol, 18(21), 2704-2711.

Liu, X., Liang, J., & Li, G. (2010). Lipopolysaccharide promotes adhesion and invasion of hepatoma cell lines HepG2 and HepG2. 2.15. Molecular biology reports, 37(5), 2235-2239.

Lu, M., Li, J., Luo, Z., Zhang, S., Xue, S., Wang, K., ... & Li, Z. (2015). Roles of dopamine receptors and their antagonist thioridazine in hepatoma metastasis. OncoTargets and therapy, 8, 1543.

Lucky, A. W., Piacquadio, D. J., Ditre, C. M., Dunlap, F., Kantor, I., Pandya, A. G., ... & Tharp, M. D. (2004). A randomized, placebo-controlled trial of 5% and 2% topical minoxidil solutions in the treatment of female pattern hair loss. Journal of the American Academy of Dermatology, 50(4), 541-553.

Ma, R., Zhang, W., Tang, K., Zhang, H., Zhang, Y., Li, D., ... & Ji, T. (2013). Switch of glycolysis to gluconeogenesis by dexamethasone for treatment of hepatocarcinoma. Nature communications, 4.

Ma, Y., Wang, J., Liu, L., Zhu, H., Chen, X., & Pan, S., et al. (2011). Genistein potentiates the effect of arsenic trioxide against human hepatocellular carcinoma: role of akt and nuclear factor-κb. Cancer Letters, 301(1), 75-84.

McKillop, I. H., & Schrum, L. W. (2009, May). Role of alcohol in liver carcinogenesis. In Seminars in liver disease (Vol. 29, No. 02, pp. 222-232). © Thieme Medical Publishers.

Merck. (1989). Merck Index: An Encyclopedia of Chemicals, Drugs and Biologicals. Merck.

Parija, S. C. (2014). Textbook of microbiology & immunology. Elsevier Health Sciences.

Sanseau, P., Agarwal, P., Barnes, M. R., Pastinen, T., Richards, J. B., Cardon, L. R., & Mooser, V. (2012). Use of genome-wide association studies for drug repositioning. Nature biotechnology, 30(4), 317-320.

Schumacher, G., Oidtmann, M., Rosewicz, S., Langrehr, J. M., Jonas, S., Mueller, A. R., ... & Gerlach, H. (2002, August). Sirolimus inhibits growth of human hepatoma cells in contrast to tacrolimus which promotes cell growth. In Transplantation proceedings (Vol. 34, No. 5, pp. 1392-1393). Elsevier.

Shang, F., Liu, M., Li, B., Zhang, X., Sheng, Y., & Liu, S., et al. (2016). The anti-angiogenic effect of dexamethasone in a murine hepatocellular carcinoma model by augmentation of gluconeogenesis pathway in malignant cells. Cancer Chemotherapy and Pharmacology, 77(5), 1-10.

Smyth, G. K. (2004). Linear models and empirical bayes methods for assessing differential expression in microarray experiments. Stat Appl Genet Mol Biol, 3(1), 3.

Tseng, P. L., Tai, M. H., Huang, C. C., Wang, C. C., Lin, J. W., & Hung, C. H., et al. (2008). Overexpression of vegf is associated with positive p53 immunostaining in hepatocellular carcinoma (hcc) and adverse outcome of hcc patients. Journal of Surgical Oncology, 98(5), 349.

Walter, E. D. (1941). Genistin (an isoflavone glucoside) and its aglucone, genistein, from soybeans. Journal of the American chemical Society, 63(12), 3273-3276.

Wang, S. D., Chen, B. C., Kao, S. T., Liu, C. J., & Yeh, C. C. (2014). Genistein inhibits tumor invasion by suppressing multiple signal transduction pathways in human hepatocellular carcinoma cells. BMC complementary and alternative medicine, 14(1), 26.

Weiss, R. B. (1992). The anthracyclines: will we ever find a better doxorubicin?. Seminars in Oncology, 19(6), 670.

Yan, H., Wang, S., Yu, H., Zhu, J., & Chen, C. (2013). Molecular pathways and functional analysis of mirna expression associated with paclitaxel-induced apoptosis in hepatocellular carcinoma cells.Pharmacology, 92(3-4), 167.

Yang, S., Guo, R., Huang, L., Yang, L., & Jiang, D. (2012). Nimesulide inhibits the proliferation of HepG2 by up-regulation of Smad4. Indian journal of pharmacology, 44(5), 599.

Ye, H., Yang, L., Cao, Z., Tang, K., & Li, Y. (2012). A pathway profile-based method for drug repositioning. Chinese Science Bulletin, 57(17), 2106-2112.

Yu, J., Qiao, L., Zimmermann, L., Ebert, M. P., Zhang, H., & Lin, W., et al. (2006). Troglitazone inhibits tumor growth in hepatocellular carcinoma in vitro and in vivo. Digest of the World Core Medical Journals, 43(1), 134–143.

Zhang, T., Hu, Y., Wu, X., Ma, R., Jiang, Q., & Wang, Y. (2016). Identifying Liver Cancer-Related Enhancer SNPs by Integrating GWAS and Histone Modification ChIP-seq Data. BioMed Research International, 2016.

Zhao, H., Jin, G., Cui, K., Ren, D., Liu, T., Chen, P., ... & Chang, J. (2013). Novel modeling of cancer cell signaling pathways enables systematic drug repositioning for distinct breast cancer metastases. Cancer research, 73(20), 6149-6163.